\documentclass[12pt]{article}
\usepackage{amsmath,amsfonts,amssymb,amsthm,textcomp}
\usepackage{amsthm}
\usepackage{hyperref}
\usepackage[font=footnotesize,labelsep=period]{caption}
\usepackage{float}
\usepackage{multirow}
\usepackage{multicol}
\usepackage{graphicx}

\def\v #1{\vert #1\vert}             

\def\n #1 #2{(-1)^{{\v #1} {\v #2}}} 

\theoremstyle{plain}

\theoremstyle{definition}

\def\<#1>{\langle#1\rangle}

\begin{document}

\centerline{\Large \bf Classical Lie symmetries and reductions for a generalized}\vskip 0.35cm \centerline{\Large \bf NLS equation in $2+1$ dimensions}\vskip 0.35cm
\centerline{\Large \bf } \vskip 0.7cm

\centerline{P. Albares$^{a)}$, J.M. Conde$^{b)}$ and P.G. Est\'evez$^{a)}$ }
\vskip 0.5cm
\centerline{$^{a)}$ Departamento de F\'isica Fundamental, Universidad de
Salamanca, Spain}\vskip 0.5cm
\centerline{$^{b)}$ Universidad San Francisco de Quito (USFQ), Quito, Ecuador}
\centerline{Departamento de Matem\'aticas, Colegio de Ciencias e Ingenierias}

\vskip 1cm

\begin{abstract}
 \noindent A non-isospectral linear problem for an integrable $2+1$ generalization of the non linear Schr\"odinger equation, which includes dispersive terms of third and fourth order, is presented. The classical symmetries of the Lax pair and the related reductions are carefully studied. We obtain several reductions of the Lax pair that yield in some cases non-isospectral problems in $1+1$ dimensions.
\end{abstract}



\section{Introduction}
Recently, an integrable system \cite{estevez16} in $2+1$ dimensions has been proposed as  a promising starting 
model for describing energy transfer processes in $\alpha$-helical proteins within the continuum limit. This model is described  by the following system of equations:
\begin{eqnarray}
&&iu_t+u_{xy}+2um_y+i\gamma_2\left(u_{xxx}-6u\omega u_x\right)
\nonumber\\
&&\quad\quad+\gamma_1 (u_{xxxx}-8u\omega u_{xx}-2u^2\omega_{xx}
-4uu_x\omega_x-6\omega u_x^2+6u^3\omega^2)=0\ ,\nonumber
\\
&&-i\omega_t+\omega_{xy}+2\omega m_y -i\gamma_2\left(\omega_{xxx}-6u\omega 
\omega_x\right)\nonumber\\
&&\quad\quad+\gamma_1 (\omega_{xxxx}-8u\omega \omega_{xx}-2\omega^2u_{xx}
-4\omega u_x\omega_x-6u \omega_x^2+6u^2\omega^3)=0\ ,\nonumber
\label{equation2}\\
&&(m_x+u\omega)_y=0\ , 
\label{nls}
\end{eqnarray}
where $m$ is a real field and $\omega=u^*$.
This equation can be considered as a higher order nonlinear Schr\"odinger equation (HONLS in the future) that includes third ($\gamma_2\neq 0$) and fourth ($\gamma_1\neq 0$) order derivatives with respect to $x$. A non-isospectral Lax pair for HONLS was obtained in \cite{estevez16}. It has the following form:
\begin{eqnarray}
&&{\displaystyle \psi_{{x}}=-i\lambda\,\psi-\chi\,u}, 
\nonumber\\
&&\chi_{{x}}=i\chi\,\lambda-\psi\,\omega,\nonumber\\
&&\psi_{{t}}=-i\gamma_{{2}}(-2\,i\chi\,{u}^{2}\omega-4\,i\chi\,u{\lambda}^{2}-i\psi\,u\omega_{{x}}+i\psi\,\omega u_{{x}}+2\,\psi\,u\omega\lambda+
\nonumber\\
&&\,\,\,4\,\psi\,{\lambda}^{3}+2\,\chi\,\lambda u_{{x}}\mbox{}+i\chi\,u_{{\it xx}})+i\gamma_{{1}}(-4\,i\chi\,{u}^{2}\omega\lambda-8\,i\chi\,u{\lambda}^{3}-2\,i\psi\,u\lambda\omega_{{x}}+
\nonumber\\
&&\,\,\, 2\,i\psi\,\omega\lambda u_{{x}}+3\,\psi\,{u}^{2}{\omega}^{2}\mbox{}+4\,\psi\,u\omega{\lambda}^{2}+8\,\psi\,{\lambda}^{4}+2\,i\chi\,\lambda u_{{\it xx}}+6\,\chi\,u\omega u_{{x}}+
\nonumber\\
&&\,\,\, 4\,\chi\,{\lambda}^{2}u_{{x}}-\psi\,u\omega_{{\it xx}}-\psi\,\omega u_{{\it xx}}+\psi\,u_{{x}}\omega_{{x}}\mbox{}-\chi\,u_{{\it xxx}})+
2\,\lambda\psi_{{y}}+i\psi\,m_{{y}}-i\chi\,u_{{y}}+\psi\lambda_y,
\nonumber\\
&&\chi_{{t}}=i\gamma_{{2}}(2\,i\psi\,u{\omega}^{2}+4\,i\psi\,\omega{\lambda}^{2}+2\,\chi\,u\omega\lambda+4\,\chi\,{\lambda}^{3}-i\psi\,\omega_{{\it xx}}-
\nonumber\\
&&\,\,\, i\chi\,u\omega_{{x}}\mbox{}+i\chi\,\omega u_{{x}}+2\,\psi\,\lambda \omega_{{x}})+i\gamma_{{1}}(-4\,i\psi\,u{\omega}^{2}\lambda -8\,i\psi\,\omega{\lambda}^{3}+2\,i\chi\,u\lambda \omega_{{x}}-
\nonumber\\
&&\,\,\, 2\,i\chi\,\omega\lambda u_{{x}}-3\,\chi\,{u}^{2}{\omega}^{2}\mbox{}-4\,\chi\,u\omega{\lambda}^{2}-8\,\chi\,{\lambda}^{4}+2\,i\psi\,\lambda \omega_{{\it xx}}-6\,\psi\,u\omega\omega_{{x}}-
\nonumber\\
&&\,\,\, 4\,\psi\,{\lambda}^{2}\omega_{{x}}+\chi\,u\omega_{{\it xx}}+\chi\,\omega u_{{\it xx}}-\chi\,u_{{x}}\omega_{{x}}\mbox{}+\psi\,\omega_{{\it xxx}})+2\,\lambda\psi_{{y}}-i\psi\,m_{{y}}+i\chi\,\omega_{{y}}+\chi\lambda_y,\qquad\quad
\label{lax}
\end{eqnarray}

where $\psi(x,y,t)$ and $\chi(x,y,t)$ are the eigenfunctions,  $i$ is the imaginary unit ($i^2=-1$),  and   $\lambda(y,t)$ is the spectral parameter. Furthermore, the equations for $\psi$ are the complex conjugate of the equations for $\chi$.
The compatibility condition of the cross-derivatives yields (\ref{nls}) as well as the non-isospectral condition
\begin{equation}\lambda_t-2\lambda\lambda_y=0.\end{equation}
The system (\ref{nls}) generalizes to $2+1$ dimensions the system proposed by Ankiewitz {\it et al} in \cite{ank}. This equation of reference  \cite{ank} contains many integrable particular cases,
such as the standard NLS equation $( \gamma_1= \gamma_2=0)$, the Hirota
equation $ (\gamma_1
 = 0)$ \cite{hirota} and the Lakshmanan-Porsezian-
Daniel equation $(\gamma_2 = 0)$ \cite{laks}. Furthermore (\ref{nls}), when $\gamma_1= \gamma_2=0$, reduces to a $2+1$ NLS equation \cite{cal} that has been extensively analyzed in \cite{eh}.

The classical Lie approach is a very well established procedure \cite{nucci}, \cite{olver} to get point symmetries of a system of differential equations. Nevertheless, in this paper we are more concerned with the identification of symmetries of the Lax pair (\ref{lax}). This approach \cite{EGP}, \cite{leg} has the benefit that the reduction associated to each symmetry of the Lax pair provides, not only the reduction of the fields $u$, $\omega$ and $m$, but the reductions of the eigenfuntions and the spectral parameter itself \cite{estevez13}, \cite{estevez17}.

We shall apply, in section 2, the classical Lie procedure to identify the symmetries of (\ref{lax}) in the four cases that arise from the different combinations of the possible values of $\gamma_1$ and $\gamma_2$.
Sections 3, 4, 5 and 6 are devoted to consider the reductions associated to the symmetries identified in section 2 for the above mentioned four different cases. We close with a section of conclusions.

\section{Classical Lie symmetries}
 
In this section we apply the classical Lie approach \cite{lie}, \cite{ovs}, in order to obtain symmetries of (\ref{lax}).
Let us consider the following infinitesimal transformation \cite{stephani}

\begin{eqnarray}
&&x\rightarrow x+\varepsilon \xi_1(x,y,t,u,\omega,m,\lambda,\psi,\chi),
\nonumber\\
&&
y\rightarrow y+\varepsilon \xi_2(x,y,t,u,\omega,m,\lambda,\psi,\chi),
\nonumber\\
&&
t\rightarrow t+\varepsilon \xi_3(x,y,t,u,\omega,m,\lambda,\psi,\chi),
\nonumber\\
&&
u\rightarrow u+\varepsilon \eta_1(x,y,t,u,\omega,m,\lambda,\psi,\chi),
\nonumber\\
&&
\omega\rightarrow \omega+\varepsilon \eta_2(x,y,t,u,\omega,m,\lambda,\psi,\chi),
\nonumber\\
&&
m\rightarrow m+\varepsilon \eta_3(x,y,t,u,\omega,m,\lambda,\psi,\chi),
\nonumber\\
&&
\lambda \rightarrow \lambda+\varepsilon \eta_4(x,y,t,u,\omega,m,\lambda,\psi,\chi),
\nonumber\\
&&
\psi \rightarrow \psi +\varepsilon \phi_1(x,y,t,u,\omega,m,\lambda,\psi,\chi),
\nonumber\\
&&
\chi \rightarrow \chi +\varepsilon \phi_2(x,y,t,u,\omega,m,\lambda,\psi,\chi),
\label{infinitesimal}
\end{eqnarray}
where $\varepsilon$ is the parameter of the Lie group and $\xi_1$, $\xi_2$, $\xi_3$, $\eta_1$, $\eta_2$, $\eta_3$, $\eta_4$, $\phi_1$ and $\phi_2$  are the infinitesimals of the vector field

\begin{equation}\label{car}
X = \xi_1 \frac{\partial}{\partial x} + 
\xi_2 \frac{\partial}{\partial y}+
\xi_3 \frac{\partial}{\partial t}+
\eta_1 \frac{\partial}{\partial u}+
\eta_2 \frac{\partial}{\partial \omega}+
\eta_3 \frac{\partial}{\partial m}+
\eta_4 \frac{\partial}{\partial \lambda}+
\phi_1 \frac{\partial}{\partial \psi }+
\phi_2 \frac{\partial}{\partial \chi }.
\end{equation}
This infinitesimal transformation induces a well known  one in the derivatives of the fields \cite{olver} \cite{stephani}. This procedure, when applied to  (\ref{lax}), yields an overdetermined system o PDEs, whose solution provides the infinitesimals. 
We shall remark that Lie approach requires a substitution of the higher order derivatives. The order of the higher derivatives in \eqref{lax}  is different depending whether $\gamma_1$ and $\gamma_2$ are (one or both) equal or different from 0. It means that it  is necessary to split the problem in four different cases depending on the different combinations of $\gamma_1$ and $\gamma_2$. 

As we said in the introduction, we are dealing with symmetries and reductions  of (\ref{lax}). These symmetries obviously provide the corresponding symmetries and reductions for (\ref{nls}). When  the infinitesimals (\ref{infinitesimal}) are known, we can proceed to calculate  the associated reductions by solving the following characteristic equation 

\begin{equation} \label{characteristic}
\frac{dx}{\xi_{1}}=\frac{dy}{\xi_{2}}=\frac{dt}{\xi_{3}}=\frac{du}{\eta_{1}}=\frac{d\omega}{\eta_{2}}=\frac{dm}{\eta_{3}}=\frac{d\lambda}{\eta_{4}}=\frac{d\psi}{\phi_{1}}=\frac{d\chi}{\phi_{2}}.
\end{equation}

\subsection{The case $\gamma_1\neq 0$ and $\gamma_2\neq 0$}

By applying the infinitesimal transformation \eqref{infinitesimal} to \eqref{lax} and following the standard procedure \cite{olver},\cite{stephani}, we obtain the  infinitesimals listed in Table 1.
\begin{table}[H]
\centering
\resizebox{\textwidth}{!} {
\begin{tabular}{|c|c|c|c|c|c|}
\hline
 \multicolumn{2}{|p{0.33\textwidth}|}{\centering ${\displaystyle \xi_{{1}}=K_{{1}}\left(t\right)}$} & \multicolumn{2}{|p{0.33\textwidth}|}{\centering ${\displaystyle \xi_{{2}}=\alpha_{{1}}}$} & \multicolumn{2}{|p{0.33\textwidth}|}{\centering ${\displaystyle \xi_{{3}}=\alpha_{{2}}}$}\\ \hline\hline
\multicolumn{3}{|p{0.5\textwidth}|}{\centering ${\displaystyle \eta_{{1}}=iu\left(\dot K_1(t)y+2\,K_{{2}}\left(t\right)\right)}$} & \multicolumn{3}{|p{0.5\textwidth}|}{\centering ${\displaystyle \eta_{{2}}=-iw\left(\dot K_1(t)y+2\,K_{{2}}\left(t\right)\right)}$}\\ \hline
\multicolumn{3}{|p{0.5\textwidth}|}{\centering ${\displaystyle \eta_{{3}}=\frac{1}{4}\ddot K_1(t)y^2+\dot K_2(t)y+K_3(x,t)}$} 
& \multicolumn{3}{|p{0.5\textwidth}|}{\centering ${\displaystyle \eta_{{4}}=0}$}\\ \hline\hline
\multicolumn{3}{|p{0.5\textwidth}|}{\centering ${\displaystyle \phi_{{1}}=\psi\left(\frac{i}{2}\dot K_1(t)y+iK_{{2}}\left(t\right)+K_{{0}}\left(y,t,\lambda\right)\right)}$} & \multicolumn{3}{|p{0.5\textwidth}|}{\centering ${\displaystyle \phi_{{2}}=\chi\left(-\frac{i}{2}\dot K_1(t)y-iK_{{2}}\left(t\right)+K_{{0}}\left(y,t,\lambda\right)\right)}$}\\ \hline
\end{tabular}
}
\label{Tab1}
\caption{$\gamma_1\neq 0$, $\gamma_2\neq 0$}
\end{table}

The dot denotes derivative with respect to $t$. $\alpha_1$ and $\alpha_2$ are arbitrary constants and $K_i$ ($i=1,...,3$) are arbitrary  functions of the indicated variables. Furthermore $K_0(y,t,\lambda)$ should satisfy the equation 

\begin{equation}
{\frac{\partial K_{{0}}\left(y,t,\lambda\right)}{\partial t}}-2\,\lambda\,{\frac{\partial K_{{0}}\left(y,t,\lambda\right)}{\partial y}}=0.\label{k0}
\end{equation}

\subsection{The case $\gamma_1\neq 0$ and $\gamma_2= 0$}
In this section we are considering \eqref{lax}, when $\gamma_2=0$. If we apply the infinitesimal transformation \eqref{infinitesimal}, then we obtain the following infinitesimals
\begin{table}[H]
\centering
 \resizebox{\textwidth}{!} {
\begin{tabular}{|c|c|c|c|c|c|}
\hline
 \multicolumn{2}{|p{0.33\textwidth}|}{\centering ${\displaystyle \xi_{{1}}=K_{{1}}\left(t\right)}+\alpha_3\,x$} & \multicolumn{2}{|p{0.33\textwidth}|}{\centering ${\displaystyle \xi_{{2}}=\alpha_{{1}}}+3\,\alpha_3\,y$} & \multicolumn{2}{|p{0.33\textwidth}|}{\centering ${\displaystyle \xi_{{3}}=\alpha_{{2}}}+4\,\alpha_3\,t$}\\ \hline\hline
\multicolumn{3}{|p{0.5\textwidth}|}{\centering ${\displaystyle \eta_{{1}}=iu\left(\dot K_1(t)y+2\,K_{{2}}\left(t\right)\right)}-\alpha_3\,u$} & \multicolumn{3}{|p{0.5\textwidth}|}{\centering ${\displaystyle \eta_{{2}}=-i\omega\left(\dot K_1(t)y+2\,K_{{2}}\left(t\right)\right)}-\alpha_3\,\omega$}\\ \hline
\multicolumn{3}{|p{0.5\textwidth}|}{\centering ${\displaystyle \eta_{{3}}=\frac{1}{4}\ddot K_1(t)y^2+\dot K_2(t)y+K_3(x,t)}-\alpha_3\,m$} 
& \multicolumn{3}{|p{0.5\textwidth}|}{\centering ${\displaystyle \eta_{{4}}=-\alpha_3 \,\lambda}$}\\ \hline\hline
\multicolumn{3}{|p{0.5\textwidth}|}{\centering ${\displaystyle \phi_{{1}}=\psi\left(\frac{i}{2}\dot K_1(t)y+iK_{{2}}\left(t\right)+K_{{0}}\left(y,t,\lambda\right)\right)}$} & \multicolumn{3}{|p{0.5\textwidth}|}{\centering ${\displaystyle \phi_{{2}}=\chi\left(-\frac{i}{2}\dot K_1(t)y-iK_{{2}}\left(t\right)+K_{{0}}\left(y,t,\lambda\right)\right)}$}\\ \hline
\end{tabular}
}
\caption{$\gamma_1\neq 0$, $\gamma_2=0$}
\label{Tab2}
\end{table}
\noindent where $\alpha_i$ ($i=1,...,3$),  are arbitrary constants and $K_j$ ($j=1,...,3$), are arbitrary  functions of the indicated variables.  $K_0(y,t,\lambda)$ satisfies the equation (\ref{k0}).

Notice that the infinitesimals are just the same as in Table 1, except  for those terms related to the constant $\alpha_3$.

\subsection{The case $\gamma_1= 0$ and $\gamma_2\neq 0$}
The procedure \cite{olver} yields the infinitesimals listed below (Table 3).
\begin{table}[H]
\centering
\resizebox{\textwidth}{!} {
\begin{tabular}{|c|c|c|c|c|c|}
\hline
 \multicolumn{2}{|p{0.33\textwidth}|}{\centering ${\displaystyle \xi_{{1}}=K_{{1}}\left(t\right)}+\alpha_3\,x$} & \multicolumn{2}{|p{0.33\textwidth}|}{\centering ${\displaystyle \xi_{{2}}=\alpha_{{1}}}+2\,\alpha_3\,y$} & \multicolumn{2}{|p{0.33\textwidth}|}{\centering ${\displaystyle \xi_{{3}}=\alpha_{{2}}}+3\,\alpha_3\,t$}\\ \hline\hline
\multicolumn{3}{|p{0.5\textwidth}|}{\centering ${\displaystyle \eta_{{1}}=iu\left(\dot K_1(t)y+2\,K_{{2}}\left(t\right)\right)}-\alpha_3\,u$} & \multicolumn{3}{|p{0.5\textwidth}|}{\centering ${\displaystyle \eta_{{2}}=-iw\left(\dot K_1(t)y+2\,K_{{2}}\left(t\right)\right)}-\alpha_3\,\omega$}\\ \hline
\multicolumn{3}{|p{0.5\textwidth}|}{\centering ${\displaystyle \eta_{{3}}=\frac{1}{4}\ddot K_1(t)y^2+\dot K_2(t)y+K_3(x,t)}-\alpha_3\,m$} 
& \multicolumn{3}{|p{0.5\textwidth}|}{\centering ${\displaystyle \eta_{{4}}=-\alpha_3 \,\lambda}$}\\ \hline\hline
\multicolumn{3}{|p{0.5\textwidth}|}{\centering ${\displaystyle \phi_{{1}}=\psi\left(\frac{i}{2}\dot K_1(t)y+iK_{{2}}\left(t\right)+K_{{0}}\left(y,t,\lambda\right)\right)}$} & \multicolumn{3}{|p{0.5\textwidth}|}{\centering ${\displaystyle \phi_{{2}}=\chi\left(-\frac{i}{2}\dot K_1(t)y-iK_{{2}}\left(t\right)+K_{{0}}\left(y,t,\lambda\right)\right)}$}\\ \hline
\end{tabular}
}
\caption{$\gamma_1 = 0$, $\gamma_2\neq0$}
\label{Tab3}
\end{table}
\noindent $\alpha_i$ ($i=1,...,3$)  are arbitrary constants and $K_j$ ($j=1,...,3$) are arbitrary  functions of the indicated variables. $K_0(y,t,\lambda)$ satisfies the equation (\ref{k0}).

We can easily see that the only additional symmetry to those listed in Table 3 is 
the symmetry associated to $\alpha_3$.

\subsection{The case $\gamma_1= 0$ and $\gamma_2= 0$}
The infinitesimals obtained in this case are presented in Table 4.
\begin{table}[H]
\centering
\resizebox{\textwidth}{!} {
\begin{tabular}{|c|c|c|c|c|c|}
\hline
 \multicolumn{2}{|p{0.33\textwidth}|}{\centering ${\displaystyle \xi_{{1}}=K_{{1}}\left(t\right)}+\alpha_3\,x+2\alpha_5\,xt$} & \multicolumn{2}{|p{0.33\textwidth}|}{\centering ${\displaystyle \xi_{{2}}=\alpha_{{1}}}+\alpha_4\,y+2\alpha_6\,t+2\alpha_5\,yt$} & \multicolumn{2}{|p{0.33\textwidth}|}{\centering ${\displaystyle \xi_{{3}}=\alpha_{{2}}+\alpha_{{3}}t+\alpha_{{4}}t+2\alpha_{{5}}t^2}$}\\ \hline\hline
\multicolumn{3}{|p{0.5\textwidth}|}{\centering $ \eta_{{1}}=iu\left(\dot K_1(t)y+2\,K_{{2}}\left(t\right)+2x(\alpha_5 y+\alpha_6)\right)-\left(\alpha_3+2\alpha_5 t\right)u$} & \multicolumn{3}{|p{0.5\textwidth}|}{\centering ${\displaystyle \eta_{{2}}=-i\omega\left(\dot K_1(t)y+2\,K_{{2}}\left(t\right)+2x(\alpha_5 y+\alpha_6)\right)}-\left(\alpha_3+2\alpha_5 t\right)\omega$}\\ \hline
\multicolumn{3}{|p{0.5\textwidth}|}{\centering ${\displaystyle \eta_{{3}}=\frac{1}{4}\ddot K_1(t)y^2+\dot K_2(t)y+K_3(x,t)}-\left(\alpha_3 +2\alpha_5t\right)m$} 
& \multicolumn{3}{|p{0.5\textwidth}|}{\centering ${\displaystyle \eta_{{4}}=-\alpha_3 \,\lambda}-\alpha_5\left(2t\lambda+y\right)-\alpha_6$}\\ \hline\hline
\multicolumn{3}{|p{0.5\textwidth}|}{\centering ${\displaystyle \phi_{{1}}=\psi\left(\frac{i}{2}\dot K_1(t)y+iK_{{2}}\left(t\right)+K_{{0}}\left(y,t,\lambda\right)\right)}+\psi\left(ix\left(\alpha_5 y+\alpha_6\right)-\alpha_5 t\right)
$} & \multicolumn{3}{|p{0.5\textwidth}|}{\centering ${\displaystyle \phi_{{2}}=\chi\left(-\frac{i}{2}\dot K_1(t)y-iK_{{2}}\left(t\right)+K_{{0}}\left(y,t,\lambda\right)\right)}+\chi\left(-ix\left(\alpha_5 y+\alpha_6\right)-\alpha_5 t \right)$}\\ \hline
\end{tabular}
}
\caption{$\gamma_1= 0$, $\gamma_2=0$}
\label{Tab4}
\end{table}
\noindent where $\alpha_i$ ($i=1,...,6$) are arbitrary constants and $K_j$ ($j=1,...,3$), are arbitrary  functions of the indicated variables. $K_0(y,t,\lambda)$ should satisfy the equation (\ref{k0}).

In this case, we have four additional symmetries to those listed in Table 1. These new symmetries are related to the constants $\alpha_3$, $\alpha_4$, $\alpha_5$ and $\alpha_6$.

 \section{Reductions for the $\gamma_1\neq 0$ $\gamma_2\neq 0$ case}
 
In what follows, we will use the following notation: $p$ and $q$ will be the new independent variables. $\Lambda(p,q)$, $F(p,q)$, $H(p,q)$, $N(p,q)$, $\Phi(p,q,\Lambda)$ and $\Omega(p,q,\Lambda)$  are the invariants that arise from the integration of the charasteristic system (\ref{characteristic}). They correspond  to the integrations in  $\lambda$, $u$, $\omega$, $m$, $\psi$ and $\chi$. These notation can be summarized as follows:

\begin{eqnarray}
& x,y,t &\rightarrow p,q \nonumber\\
& \lambda(y,t)&\rightarrow \Lambda(p,q)\nonumber\\
& u(x,y,t) &\rightarrow F(p,q)\nonumber\\& \omega(x,y,t)&\rightarrow H(p,q)\nonumber\\& \psi(x,y,t,\lambda)&\rightarrow \Omega(p,q,\Lambda)\nonumber\\& \chi(x,y,t,\lambda)&\rightarrow \Phi(p,q,\Lambda)\nonumber\\
\end{eqnarray}

According to Table 1, we have six different reductions corresponding to two arbitrary constants and four arbitrary functions. Nevertheless, the only symmetries that yield nontrivial reductions, are those related to $K_1(t)$, $\alpha_1$ and $\alpha_2$. Therefore, we are only considering the reductions associated to these three cases. In what follows, we consider different  subcases depending which functions or constants are different or equal to zero. For instance: ``$K_1=1$" means that we are setting all functions or constants to zero with the exception of $K_1$.

\subsection{$K_1=1$}

We have to solve the characteristic system (\ref{characteristic}) that yields the following reduced variables

\begin{align}
&
p=y,
& &
{\displaystyle q=t},
\nonumber\\
&
{\displaystyle u(x,y,t)=F(p,q),}
& &{\displaystyle \omega(x,y,t)=H(p,q)},
 \nonumber\\
& 
{\displaystyle m(x,y,t)=N(p,q),}
& &
\lambda(y,t)=\Lambda(p,q),
\nonumber\\
&
{\displaystyle \psi(x,y,t,\lambda)=\Phi(p,q,\Lambda)},
& &
{\displaystyle\chi(x,y,t,\lambda)=\Omega(p,q,\Lambda)}. \label{13}
\end{align}
Nevertheless, it is not an interesting reduction because the reduced equations can be easily integrated
 providing a trivial solution
 \begin{eqnarray}
 && F(p,q)=b_0e^{iZ(p,q)},\nonumber\\
&& H(p,q)=b_0e^{-iZ(p,q)},\nonumber\\
&&N(p,q)=-3\gamma_1b_0^4p+\frac{1}{2}\int Z(p,q)_qdp\label{z}.
 \end{eqnarray}
 where $Z(p,q)$ is an arbitrary function.

\subsection{$\alpha_1=1$}
We have to solve (\ref{characteristic}) in order to have the reduction

\begin{align}
&
p=x,
& &
q=t,
\nonumber\\
&
 u(x,y,t)=F(p,q),
& &
\omega(x,y,t)=H(p,q),
\nonumber\\
&
m(x,y,t)=N(p,q),
& &
\lambda(y,t)=\Lambda(q),
\nonumber\\
&
\psi(x,y,t,\lambda)=\Phi(p,q,\Lambda),
& &
\chi(x,y,t,\lambda)=\Omega(p,q,\Lambda).
\end{align}

By introducing these new variables into (\ref{lax}), we obtain the following reduced spectral problem

\begin{eqnarray}
&&
\Phi_{{p}}+i\Lambda\,\Phi\,+F\Omega=0,
\nonumber\\
&&
\Omega_{{p}}-i\Lambda\,\Omega\,+H\Phi=0,
\nonumber\\
&&
\,\Phi_{{q}}+\gamma_1\,\{[i(-8\Lambda^4-4\Lambda^2FH-3F^2H^2+HF_{pp}+FH_{pp}-F_pH_p)+2\Lambda(F_pH-H_pF)]\Phi
\nonumber\\
&&
\quad
+[i(F_{ppp}-6FHF_p-4\Lambda^2F_p)+2\Lambda[F_{pp}-2F^2H-4\Lambda^2F]\Omega\}
\nonumber\\
&&
\quad
+\gamma_{{2}}\,\{[2\Lambda i(2\Lambda^2+FH)-(HF_p-FH_p)]\Phi+[2\Lambda iF_p-F_{pp}+2F^2H+4\Lambda^2 F]\Omega\}=0,
\nonumber\\
&&
\,\Omega_{{q}}+\gamma_1\,\{[-i(-8\Lambda^4-4\Lambda^2FH-3F^2H^2+FH_{pp}+HF_{pp}-F_pH_p)+2\Lambda(H_pF-F_pH)]\Omega
\nonumber\\
&&
\quad
+[-i(H_{ppp}-6FHH_p-4\Lambda^2H_p)+2\Lambda[H_{pp}-2H^2F-4\Lambda^2H]\Phi\}
\nonumber\\
&&
\quad
+\gamma_{{2}}\,\{[-2\Lambda i(2\Lambda^2+FH)-(HF_p-FH_p)]\Omega+[-2\Lambda iH_p-H_{pp}+2H^2F+4\Lambda^2 H]\Phi\}=0.
\nonumber\\
\label{a1}
\end{eqnarray}
whose compatibility requires $\Lambda=constant$ and provides the reduced equations

\begin{eqnarray}
&&
\left(8\,FHF_{{\it pp}}+6\,H{F_{{p}}}^{2}+2\,{F}^{2}H_{{\it pp}}+4\,FF_{{p}}H_{{p}}-F_{{\it pppp}}-6\,{F}^{3}{H}^{2}\right)\gamma_{{1}}
\nonumber\\
&&
\quad
+i\left(6\,FHF_{{p}}-F_{{\it ppp}}\right)\gamma_{{2}}\mbox{}-iF_{{q}}=0,
\nonumber\\
&&
\left(8\,FHH_{{\it pp}}+6\,F{H_{{p}}}^{2}+2\,{H}^{2}F_{{\it pp}}+4\,HF_{{p}}H_{{p}}-H_{{\it pppp}}-6\,{F}^{2}{H}^{3}\right)\gamma_{{1}}
\nonumber\\
&&
\quad
-i\left(6\,FHH_{{p}}-H_{{\it ppp}}\right)\gamma_{{2}}\mbox{}+iH_{{q}}=0
\end{eqnarray}
This reduction yields the equations and the isospectral Lax pair of reference \cite{ank}.

\subsection{$\alpha_2\neq0$}
The solution of the characteristic system (\ref{characteristic}) trivially yields the reductions
\begin{align}
&
p=x,
& &
q=y,
\nonumber\\
&
 u(x,y,t)=F(p,q),
& &
\omega(x,y,t)=H(p,q),
\nonumber\\
&
m(x,y,t)=N(p,q),
& &
\lambda(y,t)=\Lambda(q),
\nonumber\\
&
\psi(x,y,t,\lambda)=\Phi(p,q,\Lambda),
& &
\chi(x,y,t,\lambda)=\Omega(p,q,\Lambda).
\end{align}

By inserting these new variables into (\ref{lax}), we obtain the spectral problem

\begin{eqnarray}
&&
\Phi_{{p}}+i\Lambda\,\Phi\,+F\Omega=0,
\nonumber\\
&&
\Omega_{{p}}-i\Lambda\,\Omega\,+H\Phi=0,
\nonumber\\
&&
2\,\Lambda\,\Phi_{{q}}=i\,F_{{q}}\Omega-i\,N_{{q}}\Phi+\gamma_1\,\{[i(-8\Lambda^4-4\Lambda^2FH-3F^2H^2+HF_{pp}+FH_{pp}-F_pH_p)
\nonumber\\
&&
\quad
+2\Lambda(F_pH-H_pF)]\Phi
+[i(F_{ppp}-6FHF_p-4\Lambda^2F_p)+2\Lambda[F_{pp}-2F^2H-4\Lambda^2F]\Omega\}
\nonumber\\
&&
\quad
+\gamma_{{2}}\,\{[2\Lambda i(2\Lambda^2+FH)-(HF_p-FH_p)]\Phi+[2\Lambda iF_p-F_{pp}+2F^2H+4\Lambda^2 F]\Omega\},
\nonumber\\
&&
2\,\Lambda\,\Omega_{{q}}=-i\,H_{{q}}\Phi+i\,N_{{q}}\Omega+\gamma_1\,\{[-i(-8\Lambda^4-4\Lambda^2FH-3F^2H^2+FH_{pp}+HF_{pp}-F_pH_p)
\nonumber\\
&&
\quad
+2\Lambda(H_pF-F_pH)]\Omega
+[-i(H_{ppp}-6FHH_p-4\Lambda^2H_p)+2\Lambda[H_{pp}-2H^2F-4\Lambda^2H]\Phi\}
\nonumber\\
&&
\quad
+\gamma_{{2}}\,\{[-2\Lambda i(2\Lambda^2+FH)+(HF_p-FH_p)]\Omega+[-2\Lambda iH_p-H_{pp}+2H^2F+4\Lambda^2 H]\Phi\}.
\nonumber\\
\label{a2}
\end{eqnarray}
The compatibility of (\ref{a2}) implies $\Lambda=constant$ and yields the reduced equations

\begin{eqnarray}
&&
\left(8\,FHF_{{\it pp}}+6\,H{F_{{p}}}^{2}+2\,{F}^{2}H_{{\it pp}}+4\,FF_{{p}}H_{{p}}-F_{{\it pppp}}-6\,{F}^{3}{H}^{2}\right)\gamma_{{1}}
\nonumber\\
&&
\quad
+i\left(6\,FHF_{{p}}-F_{{\it ppp}}\right)\gamma_{{2}}\mbox{}-2\,FN_{{q}}-F_{{\it pq}}=0,
\nonumber\\
&&
\left(8\,FHH_{{\it pp}}+6\,F{H_{{p}}}^{2}+2\,{H}^{2}F_{{\it pp}}+4\,HF_{{p}}H_{{p}}-H_{{\it pppp}}-6\,{F}^{2}{H}^{3}\right)\gamma_{{1}}
\nonumber\\
&&
\quad
-i\left(6\,FHH_{{p}}-H_{{\it ppp}}\right)\gamma_{{2}}\mbox{}-2\,HN_{{q}}-H_{{\it pq}}=0,
\nonumber\\
&&
FH_{{q}}+HF_{{q}}+N_{{\it pq}}=0.
\end{eqnarray}
 
 \section{Reductions for the $\gamma_1\neq 0$ $\gamma_2= 0$ case}
 
 For this section we will set $\gamma_2= 0$ in (\ref{lax}). According to Table 2, we have four non trivial reductions related to $K_1(t)$, $\alpha_1$, $\alpha_2$, and $\alpha_3$, but only the last one gives us a new result. The other three reductions provides us the same results as in previous section, by imposing  $\gamma_2=0$ in equations (\ref{z}), (\ref{a1}) and (\ref{a2}).
 
 \subsection{$\alpha_3=1$}
The solution for the characteristic equation (\ref{characteristic}) 
provides the following reduction

\begin{align}
&
p=t^{-\frac{1}{4}}\,x,
& &
q=\frac{t^{\frac{3}{4}}}{y},
\nonumber\\
&
 u(x,y,t)=t^{-\frac{1}{4}}\,F(p,q),
& &
\omega(x,y,t)=t^{-\frac{1}{4}}\,H(p,q),
\nonumber\\
&
m(x,y,t)=t^{-\frac{1}{4}}\,N(p,q),& &
\lambda(y,t)=t^{-\frac{1}{4}}\,\Lambda(q),
\nonumber\\
&
\psi(x,y,t,\lambda)=\Phi(p,q,\Lambda),& &
\chi(x,y,t,\lambda)=\Omega(p,q,\Lambda).
\end{align}

By inserting these new variables into (\ref{lax}) with $\gamma_2= 0$, we obtain the following \textbf{non-isospectral} Lax pair

\begin{eqnarray}
&&
\Phi_{{p}}+i\Lambda\,\Phi\,+F\Omega=0,
\nonumber\\
&&
\Omega_{{p}}-i\Lambda\,\Omega\,+H\Phi=0,
\nonumber\\
&&
{\displaystyle\left(\frac{8\,\Lambda \,q+3}{4}\right)q \,\Phi_q}+i\left(\frac{\Lambda\,p}{4}+q^2\,N_q\right)\Phi+q^2\Lambda_q\Phi+\left(\frac{pF}{4}-iq^2F_q\right)\Omega
\nonumber\\
&&
\quad
+\gamma_{{1}}[{\displaystyle\left\{ 2\Lambda(HF_p-FH_p)+i(-8\Lambda^4-4\Lambda^2FH+HF_{pp}+HF_{pp}-F_pH_p-3F^2H^2)\right\}\Phi}
\nonumber\\
&&
\quad
+{\displaystyle \left\{2\Lambda F_{pp}-4\Lambda F^2H-8\Lambda^3F+i(F_{ppp}-4\Lambda^2F_p-6FHF_p)\right\}}\Omega]=0,
\nonumber\\
&&
{\displaystyle\left(\frac{8\,\Lambda \,q+3}{4}\right)q \,\Omega_q}-i\left(\frac{\Lambda\,p}{4}+q^2\,N_q\right)\Omega+q^2\Lambda_q\Omega+\left(\frac{Hp}{4}+iq^2H_q\right)\Phi
\nonumber\\
&&
\quad
+\gamma_{{1}}[{\displaystyle\left\{ 2\Lambda(FH_p-HF_p)-i(-8\Lambda^4-4\Lambda^2FH+FH_{pp}+FH_{pp}-F_pH_p-3F^2H^2)\right\}\Omega}
\nonumber\\
&&
\quad
+{\displaystyle \left\{2\Lambda H_{pp}-4\Lambda H^2F-8\Lambda^3H-i(H_{ppp}-4\Lambda^2H_p-6FHH_p)\right\}}\Phi]=0.
\nonumber\\
\label{a3}
\end{eqnarray}
The compatibility of the above equations yields the reduced equations
\begin{eqnarray}
&&
\left(-6\,{F}^{3}{H}^{2}+2\,{F}^{2}H_{{\it pp}}+8\,FHF_{{\it pp}}+4\,FF_{{p}}H_{{p}}+6\,H{F_{{p}}}^{2}-F_{{\it pppp}}\right)\gamma_{{1}}
\nonumber\\
&&
\quad
+2q^2\,FN_{{q}}+{q}^{2}F_{{\it pq}}\mbox{}+\frac{i}{4}\left(pF_{{p}}-3\,qF_{{q}}+F\right)=0,
\nonumber\\
&&
\left(-6\,{F}^{2}{H}^{3}+2\,{H}^{2}F_{{\it pp}}+8\,FHH_{{\it pp}}+4\,HF_{{p}}H_{{p}}+6\,F{H_{{p}}}^{2}-H_{{\it pppp}}\right)\gamma_{{1}}
\nonumber\\
&&
\quad
+2q^2\,HN_{{q}}+{q}^{2}H_{{\it pq}}\mbox{}-\frac{i}{4}\left(pH_{{p}}-3\,qH_{{q}}+H\right)=0,
\nonumber\\
&&
FH_{{q}}+HF_{{q}}+N_{{\it pq}}=0,
\end{eqnarray}
and the\textbf{ non-isospectral} condition
\begin{equation}
\Lambda_{{q}}={\frac{\Lambda}{q\left(8\,\Lambda\,q+3\right)}}.
\end{equation}

  \section{Reductions for the $\gamma_1=0$, $\gamma_2\neq 0$ case}
  
  In this section we will set $\gamma_1= 0$ in (\ref{lax}). From Table 3, we can realize that, as in the previous section, we  have four non trivial reductions related to $K_1(t)$, $\alpha_1$, $\alpha_2$, and $\alpha_3$. The first three reductions provides  the same results of section 3 by setting  $\gamma_1=0$. 
 
 \subsection{$\alpha_3=1$}
The characteristic equation (\ref{characteristic}) yields the reduction

\begin{align}
&
p=t^{-\frac{1}{3}}x,
& &
q=\frac{t^{\frac{2}{3}}}{y},
\nonumber\\
&
 u(x,y,t)=t^{-\frac{1}{3}}F(p,q),& &
\omega(x,y,t)=t^{-\frac{1}{3}}H(p,q),
\nonumber\\
&
m(x,y,t)=t^{-\frac{1}{3}}N(p,q),& &
\lambda(y,t)=t^{-\frac{1}{3}}\Lambda(q),
\nonumber\\
& 
\psi(x,y,t,\lambda)=\Phi(p,q,\Lambda),& &
\chi(x,y,t,\lambda)=\Omega(p,q,\Lambda).
\end{align}

The introduction of these reductions into the Lax pair (\ref{lax}) yields the following  \textbf{non-isospectral }linear problem

\begin{eqnarray}
&&
\Phi_{{p}}+i\Lambda\,\Phi\,+F\Omega=0,
\nonumber\\
&&
\Omega_{{p}}-i\Lambda\,\Omega\,+H\Phi=0,
\nonumber\\
&&
\left(\frac{2}{3}\,q+2\,\Lambda\,{q}^{2}\mbox{}\right)\,\Phi_{{q}}+\frac{p}{3}\,\left(F\Omega+i\Lambda\,\Phi\right)+q^2\left(\Lambda_q\Phi+i\,N_{{q}}\Phi-i\,F_{{q}}\Omega\right)
\nonumber\\
&&
\quad
+[(2i\,FH\Lambda+4\,i{\Lambda}^{3}+FH_{{p}}-HF_{{p}})\Phi+(2\,H{F}^{2}+4\,F{\Lambda}^{2}+2\,i\Lambda\,F_{{p}}-F_{{\it pp}})\Omega]\gamma_{{2}}=0,
\nonumber\\
&&\left(\frac{2}{3}\,q+2\,\Lambda\,{q}^{2}\mbox{}\right)\,\Omega_{{q}}+\frac{p}{3}\,\left(H\Phi-i\Lambda\,\Omega\right)+q^2\left(\Lambda_q\Omega-i
\,N_{{q}}\Omega+i\,H_{{q}}\Phi\right)
\nonumber\\
&&
\quad
+[(-2i\,FH\Lambda-4\,i{\Lambda}^{3}-FH_{{p}}+HF_{{p}})\Omega+(2\,F{H}^{2}+4\,H{\Lambda}^{2}-2\,i\Lambda\,H_{{p}}-H_{{\it pp}})\Phi]\gamma_{{2}}=0.
\nonumber\\
\label{A3}
\end{eqnarray}
The compatibility of the above system provides the following equations
\begin{eqnarray}
&&
{\displaystyle \left(6\,FHF_{{p}}-F_{{\it ppp}}\right)i\gamma_{{2}}+\frac{i}{3}\left(F+pF_{{p}}-2qF_{{q}}\right)+q^2\left(F_{{\it pq}}+FN_{{q}}\right)=0},
\nonumber\\
&&
{\displaystyle -\left(6\,FHH_{{p}}-H_{{\it ppp}}\right)i\gamma_{{2}}-\frac{i}{3}\left(H+pH_{{p}}-2qH_{{q}}\right)+q^2\left(H_{{\it pq}}+HN_{{q}}\right)=0},
\nonumber\\
&&
{\displaystyle FH_{{q}}+HF_{{q}}+N_{{\it pq}}=0},
\end{eqnarray}
and the\textbf{ non-isospectral} condition

\begin{equation}
\Lambda_{{q}}=\frac{1}{2}\,{\frac{\Lambda}{q\left(3\,\Lambda\,q+1\right)}}.
\end{equation}

    \section{Reductions for the $\gamma_1=0$, $\gamma_2= 0$ case}

 For this section we will set $\gamma_1= 0$ and $\gamma_2= 0$ in (\ref{lax}). We new four additional non trivial reductions
 related to the constants $\alpha_3$, $\alpha_4$, $\alpha_5$ and $\alpha_6$. 
 
  \subsection{$\alpha_3=1$}
Characteristic equation (\ref{characteristic}) yields the reductions

\begin{align}
&
p=\frac{x}{t},& &
q=y,
\nonumber\\
&
 u(x,y,t)=\frac{F(p,q)}{t},
& & \omega(x,y,t)=\frac{H(p,q)}{t},
\nonumber\\
&
m(x,y,t)=\frac{N(p,q)}{t},& &
\lambda(y,t)=\frac{\Lambda(q)}{t},
\nonumber\\
&
\psi(x,y,t,\lambda)=\Phi(p,q,\Lambda),
& &
\chi(x,y,t,\lambda)=\Omega(p,q,\Lambda).
\end{align}

By inserting these new variables into (\ref{lax}), where we have taken previously $\gamma_1=\gamma_2= 0$, we obtain

\begin{eqnarray}
&&
\Phi_{{p}}+i\Lambda\,\Phi\,+F\Omega=0,
\nonumber\\
&&
\Omega_{{p}}-i\Lambda\,\Omega\,+H\Phi=0,
\nonumber\\
&&
2\,\Lambda\,\Phi_{{q}}={\displaystyle \left(\frac{1}{2}+ip\Lambda-iN_q\right)\Phi+\left(pF+iF_q\right)\Omega},
\nonumber\\
&&
2\,\Lambda\,\Omega_{{q}}={\displaystyle \left(\frac{1}{2}-ip\Lambda+iN_q\right)\Omega+\left(pH-iH_q\right)\Phi},
\label{AA3}
\end{eqnarray}
whose compatibility provides us  $$\Lambda_q=-\frac{1}{2}$$ and the reduced equations

\begin{eqnarray}
&&
2\,FN_{{q}}+F_{{\it pq}}-ipF_{{p}}-iF=0,
\nonumber\\
&&
{\displaystyle 2\,HN_{{q}}+H_{{\it pq}}+ipH_{{p}}+iH=0},
\nonumber\\
&&
{\displaystyle FH_{{q}}+HF_{{q}}+N_{{\it pq}}=0}.
\end{eqnarray}

  \subsection{$\alpha_4=1$}
  The reductions obtained through the integration of
(\ref{characteristic}) are:

\begin{align}
&
p=x,& &
q=\frac{t}{y},
\nonumber\\
&
 u=F(p,q),& &
w=H(p,q),
\nonumber\\
&
m=N(p,q),& &
\lambda=\Lambda(q),
\nonumber\\
&\psi=\Phi(p,q),& &
\chi=\Omega(p,q).
\end{align}
The Lax pair  (\ref{lax}) reduces to the spectral problem

\begin{eqnarray}
&&
\Phi_{{p}}+i\Lambda\,\Phi\,+F\Omega=0,
\nonumber\\
&&
\Omega_{{p}}-i\Lambda\,\Omega\,+H\Phi=0,
\nonumber\\
&&
{\displaystyle \Phi_{{q}}-iq\,\Omega\,F_{{q}}+iq\,\Phi\,N_{{q}}+2\,\Lambda\,q\,\Phi_{{q}}=0},
\nonumber\\
&&
{\displaystyle \Omega_{{q}}+iq\,\Phi\,H_{{q}}-iq\,\Omega\,N_q+2\,\Lambda\,q\,\Omega_{{q}}=0}.
\label{AA3}
\end{eqnarray}

whose compatibility provides us: $\Lambda=constant$ and the reduced equations

\begin{eqnarray}
&&
{\displaystyle-iF_{{q}}+2q\,FN_{{q}}+qF_{{\it pq}}=0},
\nonumber\\
&&
{\displaystyle iH_{{q}}+2q\,HN_{{q}}+qH_{{\it pq}}=0},
\nonumber\\
&&
{\displaystyle H_{{q}}F+HF_{{q}}+N_{{\it pq}}=0}.
\end{eqnarray}

  \subsection{$\alpha_5=1$}
The integration of the characteristic system (\ref{characteristic}) yields the reduction

\begin{align}
&
{\displaystyle p={\frac{x}{t}}},& &
{\displaystyle q={\frac{y}{t}}},
\nonumber\\
&
{\displaystyle u=\frac{F(p,q)}{t}\,{\rm exp}\left(itpq\right)},
,& &
{\displaystyle w=\frac{H(p,q)}{t}\,{\rm exp}\left(-itpq\right)},
 \nonumber\\
&
{\displaystyle m={\frac{N\left(p,q\right)}{t}}}
,& &
{\displaystyle \lambda={\frac{-qt+\Lambda(p,q)}{2t}}},
\nonumber\\
&
{\displaystyle \psi=\frac{\Phi(p,q)}{\sqrt{t}}\,{\rm exp}\left(\frac{itpq}{2}\right)},& &
{\displaystyle \chi=\frac{\omega(p,q)}{\sqrt{t}}\,{\rm exp}\left(\frac{-itpq}{2}\right)}.
\end{align}

The reduced Lax pair is in this case

\begin{eqnarray}
&&
{\displaystyle \Phi_{{p}}+\frac{1}{2}\,i\,\Lambda\Phi+F\Omega=0},
\nonumber\\
&&
{\displaystyle \Omega_{{p}}-\frac{1}{2}\,i\,\Lambda\Omega+H\Phi=0},
\nonumber\\
&&
{\displaystyle \Lambda\,\Phi_{{q}}-i\,F_{{q}}\Omega+i\,N_{{q}}\Phi=0},
\nonumber\\
&&
{\displaystyle \Lambda\,\Omega_{{q}}+i\,H_{{q}}\Phi-iN_{{q}}\Omega=0}.
\label{a5}
\end{eqnarray}
whose compatibility implies that  $\Lambda=constant$ and the reduced equations

\begin{eqnarray}
&&
{\displaystyle 2\,FN_{{q}}+F_{{\it pq}}=0},
\nonumber\\
&&
{\displaystyle 2\,HN_{{q}}+H_{{\it pq}}=0},
\nonumber\\
&&
{\displaystyle FH_{{q}}+HF_{{q}}+N_{{\it pq}}=0}.
\end{eqnarray}

  \subsection{$\alpha_6=1$}
These are the reductions obtained through the integration of (\ref{characteristic})

\begin{align}
&
{\displaystyle p=x},
&&
{\displaystyle q=t},
\nonumber\\
&
{\displaystyle u=F(p,q)\,{\rm exp}\left(i\,\frac{p}{q}\,y\right)},
&&{\displaystyle \omega=H(p,q)\,{\rm exp}\left(-i\,\frac{p}{q}\,y\right)},
 \nonumber\\
&
{\displaystyle m=N(p,q)},
&&
{\displaystyle \lambda=\frac{2\Lambda(p,q)-y}{2q}},
\nonumber\\&
{\displaystyle \psi=\Phi(p,q)\,{\rm exp}\left(i\,\frac{p}{2q}\,y\right)},&&
{\displaystyle \chi=\Omega(p,q)\,{\rm exp}\left(-i\,\frac{p}{2q}\,y\right)}.
\end{align}
The reduced Lax pair is

\begin{eqnarray}
&&
{\displaystyle q\,\Phi_{{p}}+q\,F\,\Omega+i\Lambda\,\Phi\,=0},
\nonumber\\
&&
{\displaystyle q\,\Omega_{{p}}+q\, H\,\Phi-i\Lambda\,\Omega=0},
\nonumber\\
&&
{\displaystyle -2\,{q}^{2}\Phi_{{q}}+ 2\,pq\,F\Omega+2ip\,\Lambda\,\Phi-q\,\Phi=0},
\nonumber\\
&&
{\displaystyle -2\,{q}^{2}\Omega_{{q}}+2\,pq\,H\Phi-2ip\,\Lambda\,\Omega-q\,\Omega=0},
\label{a6}
\end{eqnarray}
whose compatibility implies $\Lambda=constant$. The reduced equations are

\begin{eqnarray}
&&
{\displaystyle pF_{{p}}+qF_{{q}}+F=0},
\nonumber\\
&&
{\displaystyle pH_{{p}}+qH_{{q}}+H=0}.
\nonumber\\
\end{eqnarray}
\section{Conclusions}
We have determined the classical Lie symmetries of a rather complicated non-isospectral Lax pair in $2+1$ dimensions, which generalizes the well known nonlinear Schr\"odinger equation. This procedure allows us to get the infinitesimals related to the independent variables and fields, as well as those associated to the spectral parameter and eigenfunctions. Four different sets of symmetries can be obtained depending whether the parameters $\gamma_1, \gamma_2$ are zero or different from zero.

The next step is the identification of the reductions associated to each symmetry. Our procedure has the advantage that the reduced spectral parameter and eigenfunctions are simultaneously determined. Therefore, we can have, not only the reduced equations, but the reduced spectral problem. Actually, three special cases have been obtained, in which the reduced spectral problem in $1+1$ dimensions is yet non-isospectral.

We should remark that the reduced equations are, in most of the cases, quite complicated and, in many cases, non-autonomous. The identification of the integrability of these equations and its associated Lax pair could be in general a non easy problem. Nevertheless, the identification of the symmetries of (\ref{lax}) yields the right reduced spectral problem.

\section*{Acknowledgments}
 This research has been supported by MINECO (Grants MAT2013-46308 and MAT2016-75955) and Junta de Castilla y Le\'{o}n (Grant SA226U13). P. Albares acknowledges a fellowship from the Junta de Castilla y Leon.


\begin{thebibliography}{00}
\bibitem{ank} A. Ankiewicz, Y. Wang, S. Wabnitz, and N. Akhmediev, Extended nonlinear Schr\"{o}dinger equation with higher-order odd and even terms and its rogue wave solutions,
{\it Phys. Rev. E}, {\bf 89} (2014) 012907 .

\bibitem{cal} F. Calogero, A Method to Generate Solvable Nonlinear Evolution Equations, {\it Lett. Nuovo Cimento}, \textbf{14} (1975) 443.

\bibitem{eh} P. G. Estev\'ez and G. A. Hernaez. Painlev\'e Analysis and Singular Manifold Method fort a 2+1 Dimensional Nonlinear Schr\"{o}dinger equation, {\it J. Nonlinear Math.
Phys.}, \textbf{8} (2001) 106.

\bibitem{EGP}
P.G. Est\'evez, M.L. Gandarias and J. Prada, Symmetry reductions of a  2+1 Lax pair,
{\it Phys. Lett. A}, {\bf 343}  (2005) 40-47.

    \bibitem{estevez13} P. G. Est\'evez · J. D. Lejarreta and C. Sard\'on, Integrable 1+1 dimensional hierarchies arising from reduction of a non-isopectral problem in $2+1$ dimensions, {\it Applied Mathematics and Computation}, \textbf{224} (2013) 311-324.
    
           \bibitem{estevez16}  P.G. Est\'evez, E. Diaz, F. Dominguez-Adame, Jose M. Cervero and E. Diez, Lump solitons in a higher-order nonlinear equation in 2+1 dimensions, {\it Phys. Rev. E}, \textbf{93} (2016) 062219.
       
           
\bibitem{estevez17} P. G. Est\'evez · J. D. Lejarreta and C. Sard\'on, Symmetry computation and reduction
of a wave model in $2 + 1$ dimensions, {\it Nonlinear Dyn}, \textbf{87} (2017) 13-23.

\bibitem{hirota} R. Hirota, Exact envelope-soliton solutions of a nonlinear wave equation, {\it J. Math. Phys.}, \textbf{14}  (1973) 805.

\bibitem{nucci}
M. C. Nucci,
{ The role of symmetries in solving differential equations},
{\it Math. Comput. Modelling}, {\bf 25} (1997) 181--193.

\bibitem{laks} M. Lakshmanan, K. Porsezian, and M. Daniel, Effect of discreteness on the continuum limit of the Heisenberg spin chain, {\it Phys. Lett. A}, \textbf{ 133}  (1988) 483-488.

\bibitem{leg}
M. Legare, Symmetry Reductions of the Lax Pair of the Four-Dimensional Euclidean Self-Dual Yang-Mills Equations,
{\sl J. Nonlinear Math. Phys.}, {\bf 3} (1996) 266-285.

\bibitem{lie}
S. Lie, {\it Theorie der Transformationgruppen}. (Teubner, Leipzig, 1988, 1890, 1893).

\bibitem{olver}
P.J. Olver, 
{\it Applications of Lie Groups to Differential Equations}. (Springer-Verlag, 1993).

\bibitem{ovs}
L.V. Ovsiannikov,
{\it Group Analysis of Differential Equations}. (Academic Press New York 1982).

\bibitem{stephani}
H. Stephani,
{\it Differential equations: their solution using symmetries}.
 (Cambridge University Press, Cambridge, 1990).


\end{thebibliography}
\end{document}